\documentclass{article}

\usepackage[nonatbib, preprint]{neurips_2025}

\usepackage[utf8]{inputenc} 
\usepackage[T1]{fontenc}    
\usepackage{hyperref}       
\usepackage{url}            
\usepackage{booktabs}       
\usepackage{amsfonts}       
\usepackage{nicefrac}       
\usepackage{microtype}      
\usepackage{xcolor}         
\usepackage{booktabs}
\usepackage{array} 
\usepackage{tabularx}
\usepackage{makecell}

\usepackage[pdftex]{graphicx} 
\title{Assessing LLM code generation quality through path planning tasks}

\author{%
  Wanyi Chen \\
  Duke University\\
  Durham, NC 27705 \\
  \texttt{wc151@duke.edu} \\
  \And
  Meng-wen Su \\
  George Mason University \\
  Fairfax, VA 22030 \\
  \texttt{msu5@gmu.edu} \\
  \And
  Mary Cummings \\
  George Mason University \\
  Fairfax, VA 22030 \\
  \texttt{cummings@gmu.edu} \\
}

\begin{document}

\maketitle

\begin{abstract}
  As LLM-generated code grows in popularity, more evaluation is needed to assess the risks of using such tools, especially for safety-critical applications such as path planning. Existing coding benchmarks are insufficient as they do not reflect the context and complexity of safety-critical applications. To this end, we assessed six LLMs' abilities to generate the code for three different path-planning algorithms and tested them on three maps of various difficulties. Our results suggest that LLM-generated code presents serious hazards for path planning applications and should not be applied in safety-critical contexts without rigorous testing. 
\end{abstract}

\section{Introduction}


As code generated by Large Language Models (LLMs) is increasingly adopted in many domains, evaluation is needed to assess the possible risks and problems of using LLM-generated code, especially in safety-critical contexts like path planning. Path-planning algorithms play a crucial role in aircraft and vehicle navigation, especially in autonomous systems such as drones and self-driving cars. For these systems, errors in planning could lead to safety hazards and even deaths. 

To this end, we assessed six LLMs's ability to generate code for the task of generating three path-planning algorithms: Dijkstra's algorithm, the Voronoi algorithm, and the Rapidly-exploring Random Tree (RRT) algorithm. We evaluated the generated code as well as how effective the LLMs were in optimizing for the stated objectives. We then discuss the implications of the evaluation results, and hazards that result from LLM-generated code used in safety-critical contexts.
\section{Background and related work}
Most code generation evaluation benchmarks have problem sets similar to programming competitions. The problem setup typically includes a function signature and a docstring, which is natural language text that describes the purpose of the function. The LLMs are expected to automatically complete the function, and the correctness of the generated code is assessed through pre-written unit tests. HumanEval \cite{chen2021evaluating} is a widely used example of this type of benchmark. It includes 164 handwritten programming problems and an average of 7.7 unit tests per problem \cite{chen2021evaluating}. A survey by Jiang et al. \cite{jiang2024survey} provides a more comprehensive list of programming competition-style benchmarks.

However, real-world software engineering tasks are often more complex than the problems addressed in the programming competition-style benchmarks. To evaluate LLMs' ability to complete real-world software engineering tasks, some researchers proposed task-based benchmarks, such as SWE-bench \cite{jimenez2023swe}. SWE-bench utilizes 2,294 software engineering problems drawn from real GitHub issues and corresponding pull requests across 12 popular Python repositories \cite{jimenez2023swe}. Given a codebase and a description of an issue to be resolved, a language model is tasked with editing the codebase to address the issue \cite{jimenez2023swe}. However, other researchers found that some solutions were provided directly in the issue report or comments and therefore doubted whether SWE-bench could meaningfully assess LLMs' software engineering abilities \cite{aleithan2024swe}.

This experiment differed from programming competition-style benchmarks in that we did not provide function signatures. The LLMs were expected to determine the function's input and output and generate appropriate function signatures based on natural language descriptions. Furthermore, whereas most programming competition tasks focus on implementing a single function, in our experiment, the LLMs needed to generate multiple functions. The LLMs determined the number of helper functions as well as how the main function should call the helper functions. Our experiment better captured the context and complexity of real-world software engineering tasks, and further avoided the problem present in SWE-bench: no solutions were offered in our prompts.

Previous evaluations of LLM-generated code found several patterns. First, LLM code generation is non-deterministic \cite{ouyang2023llm}. LLMs could return very different code for the same prompt \cite{ouyang2023llm}. In our experiment, we accounted for the non-determinism into consideration by allowing the LLMs to regenerate responses for up to five times given the same prompt. Second, LLM-generated code is error prone. Tambon et al. identified ten distinctive bug patterns in LLM-generated code \cite{tambon2025bugs}, and Liu et al. identified five categories of hallucinations \cite{liu2024exploring}. Their evaluations were primarily based on programming competition-style questions such as HumanEval \cite{liu2024exploring, tambon2025bugs}. In our experiments, we continue to explore and evaluate patterns in LLM-generated code in safety-critical contexts.

\section{Method}
To assess LLM code generation quality, we tasked six LLMs to generate three path planning algorithms. Path planning is essentially an optimization task in that such algorithms can find the best path for the shortest distance, lowest runtime, maximum clearance around obstacles, etc. Our experiments aim to answer two research questions (RQs):
1) RQ 1: If no optimization objectives are specified, how effective is the  LLM code, and what are the default LLM objectives, and 2) RQ 2: If an optimization objective is specified, how effective is the  LLM code in optimizing for the stated objective?

\subsection{LLMs and Algorithms}

We tested six LLMs of various sizes: Mistral 7B (developed by Mistral AI, 7.3 billion parameters), Gemini (developed by Google DeepMind, 70 billion parameters), DeepSeek (developed by DeepSeek, 671 billion parameters), GPT-4o (developed by OpenAI, 1.8 trillion parameters), Copilot (developed by OpenAI, 1.8 trillion parameters, fine-tuned for code generation), and Grok 3 (developed by xAI, 2.7 trillion parameters). We prompted each LLM to generate the following path-planning algorithms, which are listed in order of increasing complexity: 
 
\begin{itemize}
    \item \textbf{Dijkstra's algorithm}: Dijkstra's algorithm is a fundamental method for finding the shortest paths from a source node to all other nodes in a graph. It uses a priority queue to track the minimum cost to reach each node, initially set to infinity, except for the source node which starts at zero. The algorithm selects the node with the lowest cost, updates its neighbors' costs, and adjusts the priority queue accordingly. This continues until the shortest paths to all nodes are determined, utilizing a systematic and greedy approach \cite{DIJKSTRA1959}. 
    \item \textbf{Voronoi algorithm}: The Voronoi diagram can be used to optimize path planning by maximizing the distance from obstacles. It partitions space into regions, forming a graph with edges that are equidistant from multiple obstacles. The Voronoi algorithm constructs a Voronoi diagram from the obstacles, removes edges that intersect with obstacles, and connects the start and goal to the nearest vertices. A shortest path algorithm, such as A\* or Dijkstra’s algorithm, is then used to find a path that travels along the edges of the Voronoi diagram. The Voronoi algorithm is especially desirable in cases where maintaining maximum clearance from obstacles is crucial \cite{Voronoi}.
    \item \textbf{Rapidly-exploring random tree (RRT)}: RRT is a key method in path finding and motion planning for robotics and autonomous vehicles. It can efficiently explore high-dimensional spaces by building a space-filling tree randomly. Starting from an initial configuration, the tree expands toward randomly chosen points while avoiding collisions and meeting movement constraints. This process continues until a path to the goal is found or a maximum iteration limit is reached. RRT is favored for its speed and effectiveness in complex environments, unlike traditional grid-based algorithms that struggle in these situations \cite{RRT}. 
\end{itemize}
\subsection{Experimental setup}

\begin{figure} 
    \centering
    \includegraphics[width=1.\linewidth]{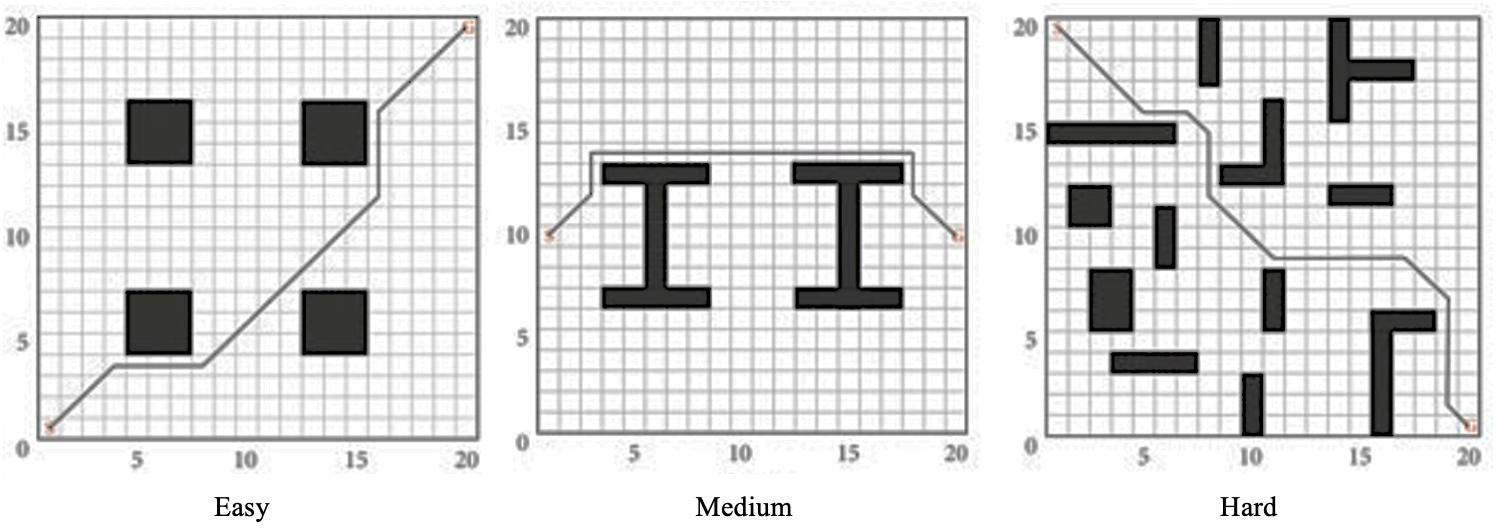}
    \caption{Increasingly difficult path planning maps}
    \label{fig:maps}
\end{figure}

Each LLM-generated algorithm was manually tested on three maps of different difficulty levels:  easy, medium, and hard, as shown in Figure \ref{fig:maps} and proposed in previous robotic path planning research \cite{maps}. The tasks described to the LLMs using the following prompts for RQ 1 and  RQ 2:

\begin{itemize}
    
    \item \textbf{RQ 1: Any-path prompt -} \textit{``Help me write a Python function code to \textbf{plan a path} on a $20 \times 20$ grid using \_\_\_\_ algorithm: \\
    1. The input of the function is an array of obstacles and the starting and ending points. The path can start and end at any (x, y) coordinates of the user's choice \\
    2. The path can include diagonal movement in addition to horizontal and vertical directions. \\
    3. The grid has rectangular obstacles defined by the coordinates of the bottom-left corner and the top-right corner. For example, “((3, 3), (5, 5))” defines an obstacle \\
    4. The function should print the grid, the obstacles, and the planned path \\
    5. You are allowed to write helper functions. \\
    6. Display the grid in an interactive plot, with x and y axes ranging from (0, 0) to (20, 20) and gridlines at every integer interval. Highlight the starting point in green and the endpoint in red. Clearly illustrate the obstacles and the planned path on the grid. \\
    7. Obstacles and the planned path should not touch; there should be a 0.5 gap between the planned path and the obstacles. \\
    8. If the goal is not reached, show the incomplete path up to the furthest point.``}

    \item \textbf{RQ 2: Shortest-path prompt -} The shortest-path prompt was the same as the any-path prompt except that in the first sentence, the phrase ``plan a path" was replaced with ``plan the shortest path."
\end{itemize}

The blank in the first sentence of the prompt was replaced with ``the Dijkstra's," ``the Voronoi," or ``the Rapidly-exploring Random Tree." We submitted the prompt to the LLMs and manually ran the generated code. If the code compiled and ran successfully, we then verified whether the code generated a viable path on at least one map. A viable path means that the path has the correct start and end points and avoids all the obstacles. We also verified that the LLMs attempted to implement the correct type of algorithm. For example, if the prompt specified ``Voronoi" but the LLM attempted to implement the Dijkstra's algorithm, the output was considered non-viable. 

\begin{figure} 
    \centering
    \includegraphics[width=1\linewidth]{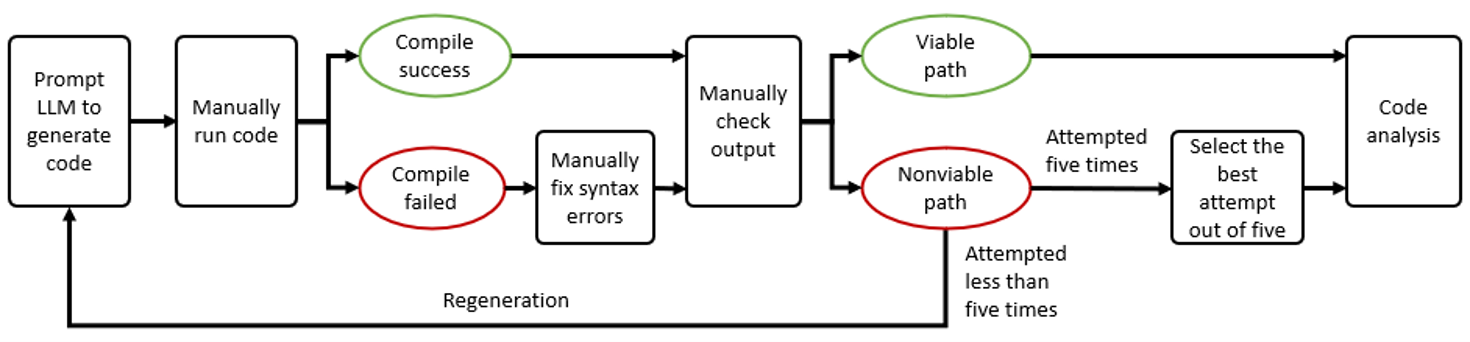}
    \caption{Experiment process diagram}
    \label{fig:diagram}
\end{figure}

If the output was non-viable, we regenerated the code by reattempting the same prompt in a new window. Each LLM was permitted up to five such attempts. If the output was still not viable after five attempts, we selected the least flawed or closest to the expected result of the five for further analysis. Figure \ref{fig:diagram} explains the process of the experiment. Once the trials were completed, we inspected the generated code including default hyperparameter choices and other implementation choices. We also assessed the following aspects of generated code:
\begin{itemize}

    \item \textit{Syntax correctness}: We checked whether the generated code contained any syntax errors, such as missing an import statement, that would prevent the code from compiling.

    \item \textit{Algorithm correctness }: We assessed whether the generated algorithm could find correct paths on all three maps. ``Correct" is defined as having the correct start and end points, not running through obstacles, and staying within the boundaries of the $20 \times 20$ grid.
    
    \item \textit{Satisfying prompt requirements}: We evaluated whether the LLM-generated code satisfied all specified requirements: whether the paths had a minimum of 0.5 distance to all obstacles and whether the grid, obstacles, and complete paths were correctly plotted. For RRT, if a path failed to reach the goal, we checked whether an incomplete path was plotted. For the Dijkstra's shortest-path prompt, we verified whether the planned paths were the shortest.
    
    \item \textit{Code Style}: For all LLM-generated code, we evaluated the code style using Pylint, an automatic code analyzer that checks for compliance with the PEP 8 Python style guide \cite{PyPI}. 
\end{itemize}

\section{Results}


The following subsections break down the results by the three algorithms.
\subsection{Dijkstra's algorithm}


In response to the RQ1 any-path prompt, the LLMs chose different default objectives, summarized in Table \ref{tab:dijkstras_default_objectives}. Gemini and Mistral generated a variation of Dijkstra's algorithm that optimized for runtime instead of the shortest path. They used the Manhattan distance as a heuristic to estimate the cost from the current node to the endpoint, which decreased the number of nodes assessed and improved overall speed. However, since heuristics provide estimates rather than precise cost values, their approach did not guarantee the shortest route. 

In contrast, GPT, Copilot, DeepSeek, and Grok attempted to implement the classical Dijkstra's algorithm that sought guaranteed shortest-path solutions through exhaustive search. They did not employ heuristics. However, while DeepSeek attempted to find the shortest path, it implemented the wrong cost calculation. It assumed a uniform cost moving from one node to every neighbor node, but the cost of moving horizontally or vertically is 1, and the cost of moving diagonally is $\sqrt{2}$. This error caused DeepSeek to generate a longer path, although it attempted to optimize for the shortest path. Appendix \ref{dijkstra} includes figures that illustrate different paths planned by different LLM implementations of Dijkstra's algorithms.


\begin{table}[h]
    \centering

    \caption{Dijkstra's default objectives and heuristics}

    \begin{tabular}{l c c c} 
        \toprule
        \textbf{LLM} & \textbf{Default objective} & \textbf{Heuristic used} & \textbf{Achieved shortest path?} \\
        \midrule
        Mistral & Runtime & Manhattan distance & No \\
        Gemini & Runtime & Manhattan distance & No \\
        DeepSeek & Shortest path & None & No (due to incorrect cost calculation) \\
        GPT-4o & Shortest path & None & Yes \\
        Copilot & Shortest path & None & Yes \\
        Grok 3 & Shortest path & None & Yes \\
        \bottomrule
    \end{tabular}

    \label{tab:dijkstras_default_objectives}
\end{table}

\begin{table}[htb]

\caption{Shortest-path prompt Dijkstra's evaluation. Green cells indicate success. The yellow cells indicate partial success or problems that are relatively easy to fix. The red cells indicate failures.} 

\includegraphics[width=1\linewidth]{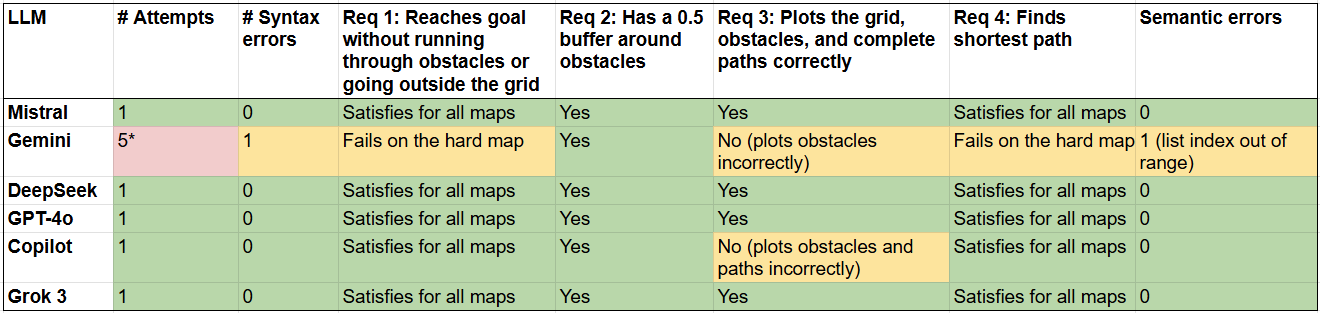}

\label{tab:dijkstra_eval_table} 
\end{table}

For the RQ2 optimization objective (shortest-path prompt), the results are shown in Table \ref{tab:dijkstra_eval_table}. The column ``\# Attempts" specifies the number of times we repeated the prompt. If an LLM could not generate viable output within five attempts, we marked it as ``5*." The column ``\# Syntax errors" specifies the number of syntax errors defined by any error that would prevent the code from compiling. If syntax errors were present, we manually fixed them before proceeding with other evaluations. The column ``Semantic errors" lists errors related to algorithm correctness.

Overall, across all maps the LLMs performed well in the task of generating the Dijkstra's algorithm. All LLMs except Gemini were successful. Gemini's code worked correctly on the easy and medium maps, but threw a list index error out of range error on the hard map, causing the program to stop before a path could be generated. In addition, Gemini and Copilot failed to generate correct plots. All other LLMs generated code that satisfied all requirements in one shot.
\subsection{Voronoi algorithm}

For both RQ1 and RQ2 prompts, none of the LLMs successfully generated a Voronoi algorithm for path generation. Table \ref{tab:voronoi_eval_table} illustrates that other than Gemini, which generated nonsensical outputs, the LLMs found paths. However, the paths were planned using either the A* or Dijkstra's algorithm rather than the Voronoi algorithm. DeepSeek and Copilot imported a Voronoi library from Scipy but never used it. Mistral, GPT 4, and Grok 3 generated incorrect Voronoi diagrams that were never used. 


\begin{table}[htb]

\caption{Shortest-path prompt Voronoi algorithm evaluation. Green cells indicate success. The yellow cells indicate partial success or problems that are relatively easy to fix. The red cells indicate failures.} 

\includegraphics[width=1.0\linewidth]{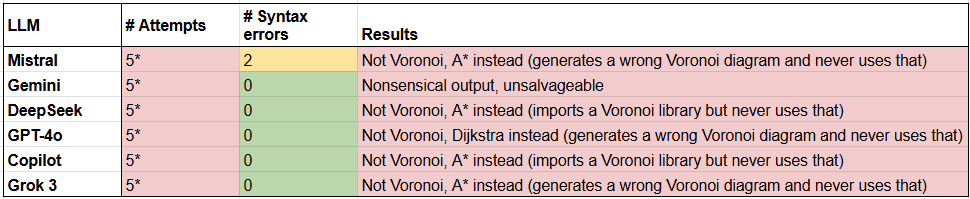}

\label{tab:voronoi_eval_table} 
\end{table}


Notably, no LLM attempted to implement the Voronoi algorithm from scratch. They all relied on the Voronoi library from Scipy. However, this library function only takes a set of points as input. The individual points in the set are considered obstacles. Since the obstacles were polygons instead of points, to properly use the library, the input set of points needed to be points along the boundaries of the polygonal obstacles. After a Voronoi diagram is generated, the Voronoi edges inside the obstacles must be removed. This is a non-trivial adaptation of the library function, and none of the LLMs implemented the correct process of constructing generalized Voronoi diagrams. 

\subsection{RRT}
For the RQ1 any-path prompt, LLMs chose different default hyperparameter values, implicitly optimizing for different objectives, listed in Appendix \ref{rrt-default}. RRT has two key hyperparameters: step size and the maximum number of iterations. The step size determines the distance the tree extends towards a sampled point in each iteration. The maximum iteration sets a limit on how many times the algorithm attempts to expand the tree before terminating. 

Different default hyperparameter values lead to different algorithm performances. A larger maximum iteration value allows the tree to expand more, covering more search space and thereby improving the success rate at the expense of increased runtime. A smaller step size allows the tree to explore more narrow passages, but also leads to slower exploration and covers less search space. It could reduce the success rate unless the maximum iteration value is set much higher. Conversely, a larger step size enables the tree to explore more quickly and covers more search space. However, if the step size is too large, the tree may bypass narrow passages, decreasing the likelihood of finding a feasible path.

Gemini and Copilot chose a smaller maximum iteration value, optimizing for runtime at the expense of success rate. Copilot (any-path prompt) further optimized for runtime by choosing a larger step size. On the other end of the runtime-success rate 
trade-off, Copilot (shortest-path prompt) did not choose a maximum iteration value. The algorithm continued searching until a path was found, no matter how many iterations. It optimized for success rate but did not consider runtime. Grok (shortest-path prompt) also optimized for success rate by choosing a very large maximum iteration value. In all other cases, the LLMs tried to balance success rate and runtime.

It is noteworthy that some LLMs did not have consistent optimization objectives. For instance, Copilot optimized for runtime for the RQ1 any-path prompt but for success rate for the RQ2 shortest-path prompt. Although the any-path and the shortest-path prompts only differed by one word, Copilot's optimization objectives differed dramatically. Similarly, Gemini and Grok 3 also had different default objectives under different prompts.

\subsubsection{RQ2: Shortest-path prompt result evaluation}

Since RRT is nondeterministic, we executed the code 100 times to determine the success rate. To assess the variance of the success rate, we executed the code 30 times and repeated this for 30 rounds. In each round, we calculated the variance of the Bernoulli distribution by $p(1-p)$, where $p$ is the success rate across the 30 executions. The average variance across the 30 rounds was reported in the parenthesis of the ``Req 1" column in Table \ref{tab:rrt_eval_table}.

Gemini had the worst performance with nonsensical output. Mistral, DeepSeek, and Copilot had similar success rates, considering the variance. GPT-4o had a slightly worse success rate as the code contained more semantic errors. Grok had the best performance, as it consistently achieved a 100\% success rate on all maps. The high success rate was due to three reasons. First, Grok correctly implemented the RRT algorithm and satisfied all prompt requirements. Second, Grok chose a very large maximum iteration value, as discussed in subsection 4.1.2. Third, Grok effectively implemented goal-biased sampling, as discussed in subsection 4.1.3.

DeepSeek, GPT-4o, and Copilot made various semantic errors. None checked the grid boundaries, so parts of their paths could land outside the grid. GPT-4o and Copilot sometimes constructed circular trees: node A's parent is node B, and node B's parent is node A. This circular tree problem caused infinite loops. GPT-4o also made an error in reconstructing incomplete paths. The program threw an error and stopped running.

\begin{table}[htb]
    \centering

    \caption{Shortest-path prompt RRT algorithm evaluation. Green cells indicate success (100\% of Requirement 1), and yellow cells indicate partial success ($70\%-99\%$ of Requirement 1) or problems that are relatively easy to fix. The red cells indicate failures ($< 70\%$ of Requirement 1).} 
    
    \includegraphics[width=1.0\linewidth]{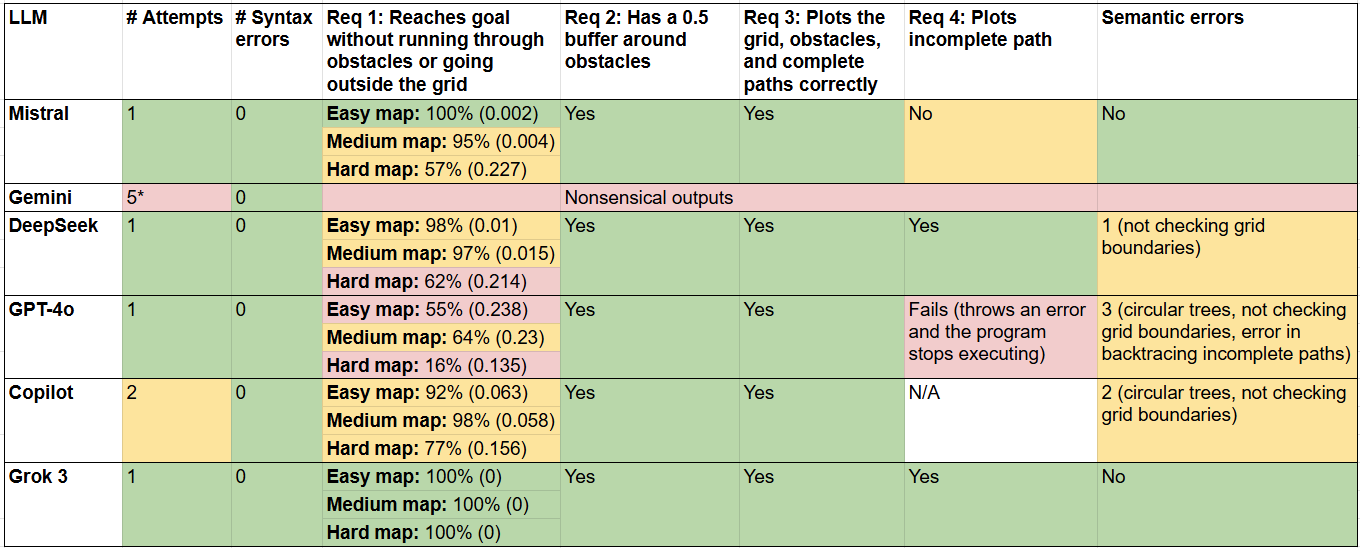}

    \label{tab:rrt_eval_table} 
\end{table}

\subsubsection{Different implementation choices}

Both Copilot (any-path prompt) and DeepSeek (both prompts) pulled a $k$-d tree library from Scipy, which theoretically can improve RRT runtime \cite{friedman1977algorithm}. However, the library had constraints, and the LLMs' failure to account for the constraints led to worse runtime. Appendix \ref{k-d-tree} includes more background information on $k$-d tree and a runtime analysis.

Another notable implementation choice is that Grok implemented goal-biased RRT for both prompts. In each iteration with a 10\% probability, instead of randomly sampling a point from the free space,  the algorithm chose the goal as the sampled point, biasing growth towards the goal instead of exploring all directions equally. Therefore, the tree reaches the goal with fewer iterations. First proposed by LaValle et al. \cite{lavalle2001randomized}, this technique improves runtime, and also results in more straight-line paths, especially near the goal. This property might be desirable in some use cases.




\begin{table}[h]
    \centering
    
    \caption{Summary of Pylint scores}

    \begin{tabularx}{\textwidth}{l c c c c >{\raggedright\arraybackslash}X}
        \toprule
        \textbf{} & \textbf{Min} & \textbf{Max} & \textbf{Average} & \textbf{Standard deviation} & \textbf{Most common issues} \\
        \midrule
        Mistral & 5.61 & 7.21 & 6.54 & 0.607 & redefined-outer-name, missing-function-docstring \\
        Gemini & 0.00 & 0.00 & 0.00 & 0.000 & bad-indentation \\
        DeepSeek & 5.82 & 7.81 & 6.55 & 0.680 & redefined-outer-name, trailing-whitespace \\
        GPT-4o & 5.00 & 8.62 & 7.30 & 1.269 & redefined-outer-name, trailing-whitespace \\
        Copilot & 3.55 & 8.38 & 6.64 & 1.675 & redefined-outer-name, trailing-whitespace \\
        Grok 3 & 5.43 & 6.85 & 6.24 & 0.582 & redefined-outer-name, trailing-whitespace \\
        \bottomrule
    \end{tabularx}    
    \label{tab:pylint_scores_width_matched}
\end{table}

\subsection{Code style evaluation}


We used Pylint to evaluate how well the LLM-generated code follows the PEP 8 Python style guide. Pylint scores range from 0 to 10, with zero indicating that the code did not meet any coding standards and ten indicating that all coding standards were met. The results are summarized in Table \ref{tab:pylint_scores_width_matched}, which reports the minimum, maximum, and average Pylint scores of all six algorithms (three for any-path prompts and three for shortest-path prompts) generated by each LLM, the standard deviation, and the most common code style issues. Appendix \ref{pylint} includes the full Pylint results.

Gemini has the worst code style. It received a Pylint score of 0 for all six algorithms, primarily for bad indentation errors. All other LLMs' average scores ranged from six to eight. Opinions vary on how Pylint scores should be interpreted: some think that Pylint scores above seven should be considered good code style \cite{yousuf2022analysis}. Judging by this standard, the LLM-generated code had roughly good code style. Others suggest that new Python files should have a Pylint score of nine or higher \cite{gramps}. Judging by this standard, none of the generated code was production-ready, and all would require manual adjustments to improve the code style. Notably, Pylint scores do not correlate with algorithm correctness. Several Voronoi implementations achieved higher than average Pylint scores. However, the generated Voronoi algorithms were far from the correct implementation.

One of the most common code style issues among multiple LLMs was ``redefined-outer-name". It happens when a variable's name hides a name defined in an outer scope or except handler \cite{Pylint_error}. The ``trailing-whitespace" error is also common. It happens when there is whitespace between the end of a line and the newline \cite{Pylint_error}. Mistral also has multiple ``missing-function-docstring" errors, which means functions do not have comments that document the purposes of the functions. These code style issues reduced the readability of the generated code and will ultimately reduce maintainability.

\section{Discussion}

The LLMs' abilities to generate code roughly corresponds to the amount of available code on Stack Overflow and GitHub. In February 2025, we searched for available Python code for each type of path-planning algorithm on both platforms, which are likely included in most LLMs' training sets. Table \ref{tab:python_code_availability} shows our search keywords and the number of search results returned. In total, Dijkstra's algorithm has the most amount of available code, and the Voronoi algorithm has the least. Correspondingly, most LLMs were most successful in implementing Dijkstra's algorithm and least successful in implementing the Voronoi algorithm. 

This relationship between code availability and LLM code generation quality suggests that LLMs do not ``understand" algorithms or coding in general. They rely heavily on existing code to generate new code. Furthermore, the LLMs do not ``understand" library documentations, as all LLMs failed to correctly use the Voronoi library. Therefore, LLM-generated code makes various mistakes and requires significant human effort to function fully. The next subsections discuss these human efforts involved in LLM code generation. 

\begin{table}[h]
    \centering
    
    \caption{The amount of available Python code (number of search results) online}

    \begin{tabular}{lccc}
        \toprule
        \textbf{Search keywords} & \textbf{Stack Overflow} & \textbf{GitHub} & \textbf{Total} \\
        \midrule
        Dijkstra’s algorithm python & 282 & 960 & 1242 \\
        Voronoi algorithm python & 55 & 87 & 142 \\
        rapidly exploring random tree algorithm python & 1 & 61 & 439 \\
        RRT algorithm python & 2 & 375 & 439 \\
        \bottomrule
    \end{tabular}
    
    \label{tab:python_code_availability}
\end{table}

\subsection{Hazard analysis: Latent dangers of using LLM-generated code}

\begin{figure}[htb]
    \centering
    \includegraphics[width=1\linewidth]{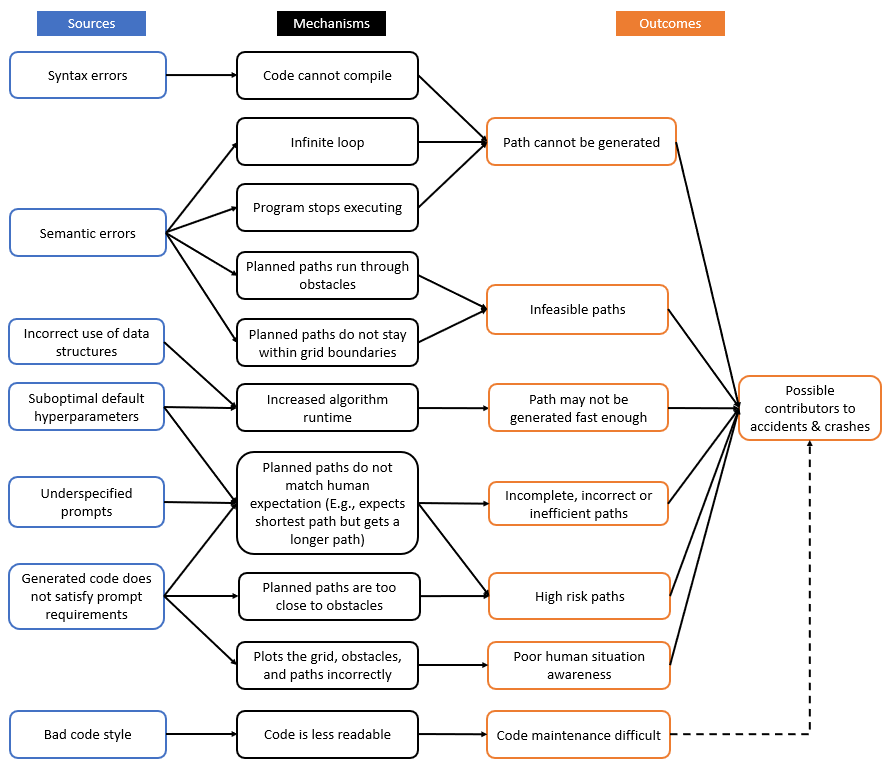}
    \caption{Hazard analysis}
    \label{fig:hazard-analysis}
\end{figure}

Path planning is a function that is fundamental to all robots, including drones, self-driving cars, and surgical robots, all of which are considered safety-critical systems. Error-prone LLM-generated path planning code in safety-critical contexts could cause delays in development, and if latent path planning errors are not caught during development and testing, the outcomes could be catastrophic. Figure \ref{fig:hazard-analysis} shows our hazard analysis following the source-mechanism-outcome framework \cite{goldberg1994system}. The sources are identified problems in LLM-generated path planning algorithms. Such sources could become possible contributors to serious problems. For example, because of semantic errors in some generated algorithms, paths could intersect with obstacles. If robots were to follow these infeasible paths, accidents and crashes would likely happen. Code problems that may seem less serious, such as bad code style, could also be possible contributors to accidents and crashes. Bad code style makes code less readable. Over time, code maintenance becomes more difficult: if engineers do not understand the existing code, they are more likely to make mistakes when integrating new code with existing code. The mistakes could eventually lead to accidents and crashes.

Much of the prior literature on AI security and LLM-related hazard analysis focuses on humans using LLMs with malicious intentions \cite{khlaaf2022hazard, shevlane2023model}. While important concerns, the dangers of over-relying on LLM-generated code in safety-critical contexts are less discussed. Harm could be caused even without malicious human intentions.

Lastly, to make LLM-generated code function properly in this effort, significant human effort was required at every step. For instance, prompt engineering is a nontrivial task. Designing prompts for code generation tasks is similar to writing software requirements specification documents, which requires both communication skills and expertise in software engineering. It is often an iterative process requiring multiple rounds of trials and errors. As seen in our experiment, slightly different prompts can lead to very different code generation. Previous research has illustrated that non-AI experts often fail in designing effective LLM prompts \cite{zamfirescu2023johnny}. Another study showed that beginning coders often struggle with writing and editing prompts to accomplish code generation tasks \cite{nguyen2024beginning}. 

Once code is generated, significant human effort is still required for testing, debugging, hyperparameter tuning, runtime optimization, and code style improvements. Appendix \ref{human-efforts} details such efforts, as well as human skills and knowledge needed for each step of this effort. Although generating code using LLMs seems automatic, this effort illustrates that human engineers cannot be replaced, especially in safety-critical applications.

\subsection{Limitations}

The utility of any hazard analysis ultimately depends on its domain of application. Problems resulting from poor path planning for small, lightweight food delivery robots will necessarily be less hazardous than those for aircraft that are fuel-constrained. Also, due to project scope constraints, we did not assess any cybersecurity issues of LLM-generated code, which could be another safety hazard. Another limitation is that as newer LLMs are developed, their ability to generate code will change, so they will need to be re-evaluated. 

\section{Conclusion}

In conclusion, we assessed six LLMs' abilities to generate the code for three different path-planning algorithms across increasingly complex scenarios. Our results show that the success of the LLM-generated code varies, with most LLMs able to successfully implement the popular Dijkstra's algorithm but none of them successfully able to generate the less-popular Voronoi algorithm. These results align with the frequency of code availability, and also indicate that LLMs do not have the ability to reason about the use of libraries. LLM-generated code also needs code style improvements.

A hazard analysis demonstrated that LLM-generated code should not be applied in safety-critical contexts without rigorous testing. Small changes in prompt wording produced very different outcomes, so reliability and code resilience is very much a concern. LLMs cannot replace human engineers for the foreseeable future, especially in safety-critical applications. 


\newpage
\bibliography{neurips_2025}


\clearpage
\onecolumn
\appendix

\section{Comparison of the LLM-generated Dijkstra's algorithm (any-path prompt)}
\label{dijkstra}
\begin{figure}[h]
    \centering
    \includegraphics[width=1\linewidth]{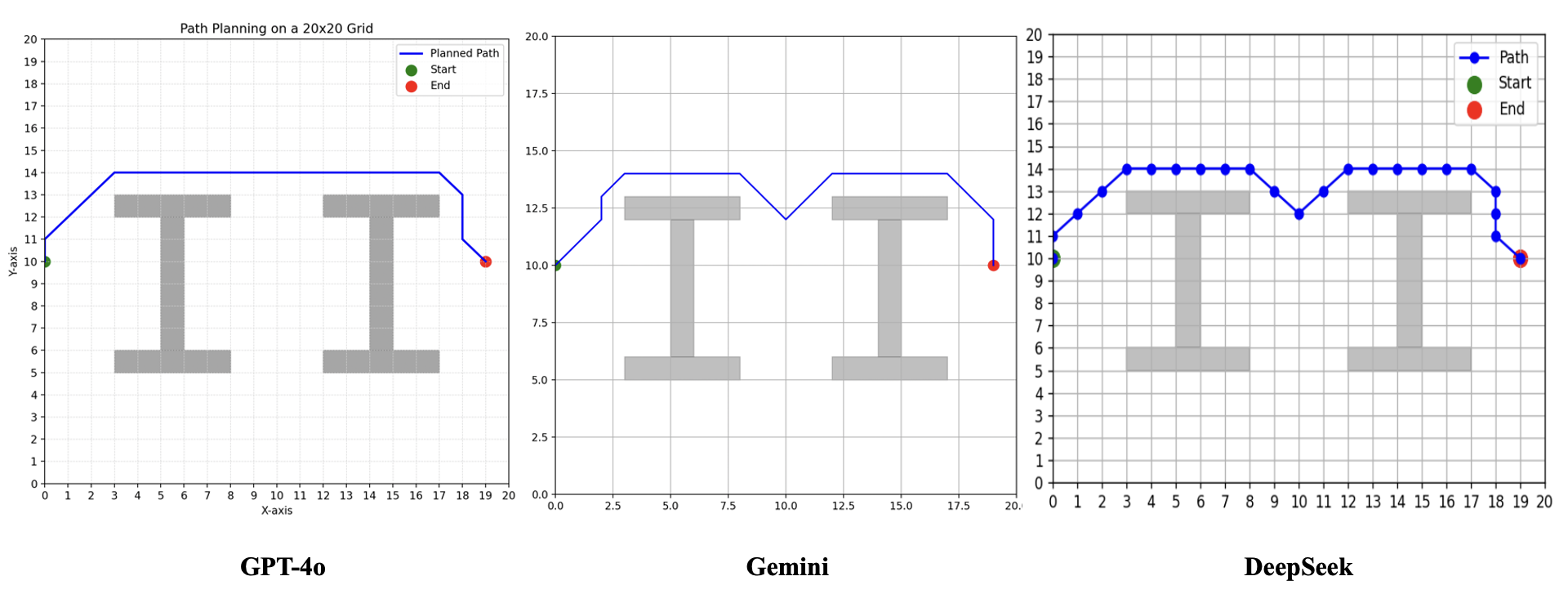}
    \caption{Performance comparison between GPT-4o, which optimized for the shortest path with correct implementation, DeepSeek, which optimized for the shortest path with incorrect implementation, and Gemini, which did not optimize for the shortest path}
    \label{fig:any_dijkstra}
\end{figure}

\clearpage

\section{RRT default hyperparameters and objectives}
\label{rrt-default}

\begin{table}[h]
    \centering
    
    \caption{RRT default hyperparameters and objectives. “Balanced” means the LLM tried to balance success rate and runtime.}

    \begin{tabularx}{\textwidth}{l c c c c c @{\hspace{0.3cm}} >{\centering\arraybackslash}X}
        \toprule
        \textbf{LLM} & \multicolumn{3}{c}{\textbf{Any-path prompt}} & \multicolumn{3}{c}{\textbf{Shortest-path prompt}} \\
        \cmidrule(lr){2-4} \cmidrule(lr){5-7}
        & \makecell{Max\\iteration} & \makecell{Step\\size} & \makecell{Default\\objective} 
        & \makecell{Max\\iteration} & \makecell{Step\\size} & \makecell{Default\\objective} \\
        \midrule
        Mistral  & 1000 & 1   & Balanced & 1000 & 1      & Balanced \\
        Gemini*  & 500  & 1   & Runtime  & 1000 & 1      & Balanced \\
        DeepSeek & 1000 & 1   & Balanced & 1000 & 1      & Balanced \\
        GPT-4o   & 1000 & 0.5 & Balanced & 1000 & 1      & Balanced \\
        Copilot  & 500 & 1.5 & Runtime & 
        \makecell[tl]{None (the algorithm\\kept running until\\ the goal was reached)} & 1 & \makecell[c]{Success rate} \\
        Grok 3   & 1000 & 1   & Balanced & 5000 & 1      & Success rate \\
        \bottomrule
    \end{tabularx}
    
    \label{tab:rrt_hyperparams_final}
\end{table}

Gemini is indicated with an asterisk (*) because, although specific step sizes and maximum iterations were chosen, it produced no output for the any-path prompt as it generated an infinite loop. In the case of the shortest-path prompt, it generated a nonsensical output. 

\clearpage

\section{Runtime analysis of $k$-d tree}
\label{k-d-tree}

For a basic RRT implementation, at each iteration, RRT generates a random point, and attempts to add a leaf node to the closest node in the direction of the random point. However, searching for the nearest neighbor of a random point is a very costly step. Taking a brute-force approach will iterate through all existing nodes, rendering a runtime complexity of $O(n)$, where $n$ is the number of existing nodes. In our experiments, all LLMs used this brute-force approach except for Copilot (any-path prompt) and DeepSeek (both prompts), which instead used a data structure called $k$-d tree \cite{atramentov2002efficient}. 

Whereas a binary search tree can effectively search through a one-dimensional array, a $k$-d tree can effectively search through a $k$ dimensional space. Our experiments plan paths on two-dimensional grids, so $k = 2$. The $k$-d tree, first proposed by Friedman et al. \cite{friedman1977algorithm}, has an average insertion cost of $O(\log n)$ and an average nearest-neighbor query cost of $O(\log n)$. Therefore, in theory, using a $k$-d tree in RRT can reduce the runtime complexity of the nearest-neighbor search step to $O(\log n)$.

Both Copilot (any-path prompt) and DeepSeek (both prompts) pulled a $k$-d tree library from Scipy (version 1.15.1), which does not have an ``insert node" function. Therefore, for every nearest-neighbor query, a $k$-d tree needs to be reconstructed from scratch. Constructing a $k$-d tree involves inserting $n$ nodes. Inserting a single node costs $O(\log n)$, so the overall runtime cost becomes $O(n \log n)$, which is worse than the brute-force cost of $O(n)$. Thus, Copilot and DeepSeek's $k$-d tree implementations relied on an external library with constraints, and failure to account for the constraints led to worse runtime.

\clearpage

\section{Pylint results}
\label{pylint}
\begin{table}[htb]
    \centering

    \caption{Pylint evaluations for different algorithms generated by the any-path prompt}

    \includegraphics[width=1\linewidth]{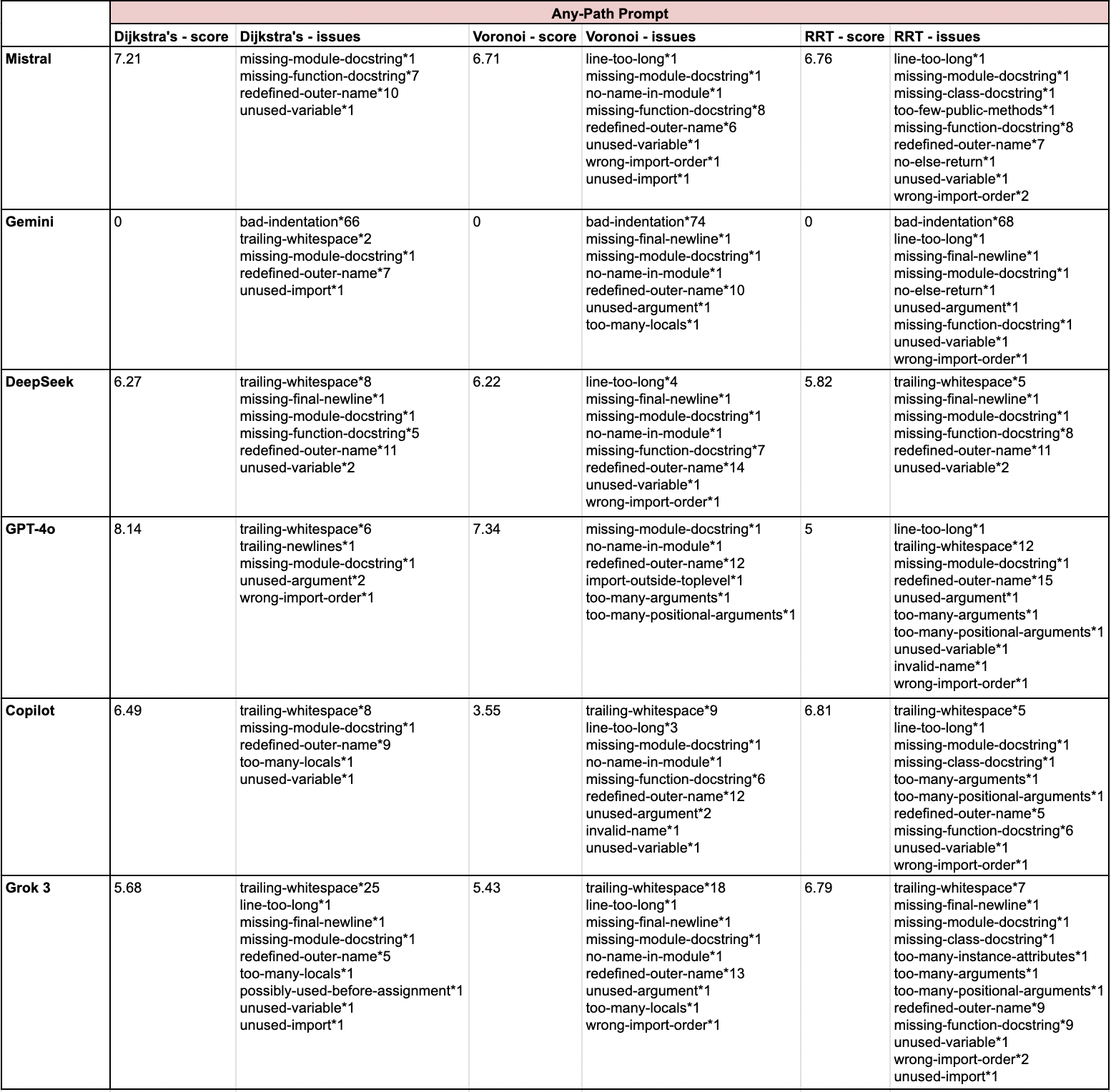}

    \label{tab:pylint_any}
\end{table}

\begin{table}[htb]
    \centering

    \caption{Pylint evaluations for different algorithms generated by the shortest-path prompt}

    \includegraphics[width=1\linewidth]{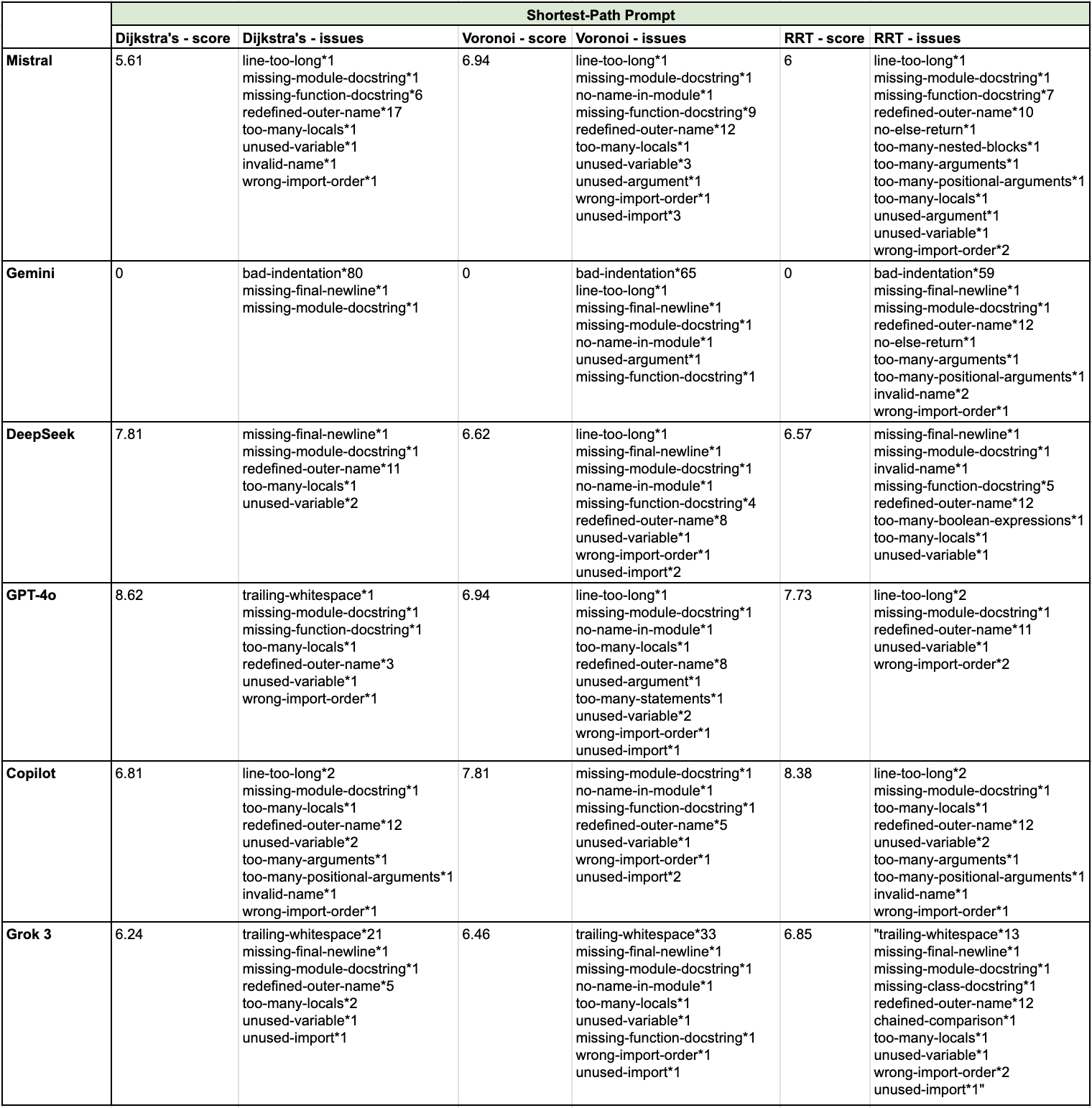}
    
    \label{tab:pylint_shortest}
\end{table}

\clearpage

\section{Human efforts in LLM code generation}
\label{human-efforts}

\begin{table}[h]
    \centering

    \caption{Human efforts involved in using LLMs for code generation}

    \begin{tabularx}{\textwidth}{>{\raggedright\arraybackslash}p{3.5cm} >{\raggedright\arraybackslash}X >{\raggedright\arraybackslash}X}
        \toprule
        \textbf{Human effort} & \textbf{Explanation} & \textbf{Human skills/knowledge needed} \\
       \midrule
       Prompt engineering and regeneration & Trials and errors are needed to design prompts that clearly specify all requirements. & Prompt engineering skills (communication skills and knowledge of LLM) and software engineering skills \\\\
        
      Testing and debugging & The generated code needs to be thoroughly tested, and bugs need to be fixed. In cases such as the Voronoi algorithm, the entire algorithm needs to be rewritten. & Software engineering skills (writing comprehensive test cases, locating bugs, and fixing bugs) and understanding of algorithms \\\\
        
        Hyperparameter tuning & If the default hyperparameters are not optimal, as seen in multiple RRT cases, the hyperparameters need to be manually tuned. & Expert understanding of algorithms (knowing the effects of each hyperparameter) \\\\
        
       Runtime optimization & Manual optimization is needed when LLMs fail to implement algorithms with efficient runtime, as seen in some LLMs’ attempts to use k-d trees in RRT. & Software engineering skills, understanding of algorithms, and expertise in runtime analysis \\\\
        
       Code style improvements & Most LLM-generated code needs manual improvements to comply with coding style guides. & Software engineering skills \\
        \bottomrule
    \end{tabularx}

    \label{tab:llm_human_effort}
\end{table}

\clearpage

\end{document}